\documentclass[superscriptaddress,preprint,prb]{revtex4-1}
\usepackage{graphicx}
\usepackage{amsmath}
\usepackage{amssymb}
\usepackage[squaren, thinqspace]{SIunits}
\usepackage{xspace}
\usepackage{setspace}
\usepackage{textcomp}
\usepackage{color}
\usepackage{soul}

\renewcommand{\degree}{\ensuremath{^\circ}\xspace}

\newcommand{\Mm}{\ensuremath{\mathbf{\overline{M}}}\xspace}

\begin{document}

\title{Magnetic microstructure and magnetotransport in Co$_2$FeAl Heusler compound thin films}
\author{Mathias Weiler}
\author{Franz D. Czeschka}
\affiliation{Walther-Mei{\ss}ner-Institut, Bayerische Akademie der Wissenschaften, 85748 Garching, Germany}
\author{Inga-Mareen Imort}
\author{G\"{u}nter Reiss}
\author{Andy Thomas}
\affiliation{Fakult\"{a}t f\"{u}r Physik, Universit\"{a}t Bielefeld, 33615 Bielefeld, Germany}
\author{Georg Woltersdorf}
\affiliation{Department of Physics, Universit\"{a}t Regensburg, 93040 Regensburg, Germany}
\author{Rudolf Gross}
\author{Sebastian T. B. Goennenwein}
\email[Electronic address: ]{goennenwein@wmi.badw.de}
\affiliation{Walther-Mei{\ss}ner-Institut, Bayerische Akademie der Wissenschaften, 85748 Garching, Germany}%

\date{\today}
\begin{abstract}
We correlate simultaneously recorded magnetotransport and spatially resolved magneto optical Kerr effect (MOKE) data in Co$_2$FeAl Heusler compound thin films micropatterned into Hall bars. Room temperature MOKE images reveal the nucleation and propagation of domains in an externally applied magnetic field and are used to extract a macrospin corresponding to the mean magnetization direction in the Hall bar. The anisotropic magnetoresistance calculated using this macrospin is in excellent agreement with magnetoresistance measurements. This suggests that the magnetotransport in Heusler compounds can be adequately simulated using simple macrospin models, while the magnetoresistance contribution due to domain walls is of negligible importance.
\end{abstract}
\maketitle
Cobalt-based Heusler compounds are an interesting class of materials for spintronic applications due to their predicted 100\% spin-polarization~\cite{Kandpal:2007} and their Curie temperature well in excess of room temperature~\cite{Trudel:2010}. Furthermore, tunneling magnetoresistance~\cite{Julliere:1975} (TMR)-ratios exceeding 1000\% have been reported~\cite{Ishikawa:2009}, making Heusler-based devices attractive for magnetic data storage applications. However, while the TMR properties have been vigorously investigated (e.g., for Co$_2$FeAl~\cite{Ebke:2009, Inomata:2006, Hirohata:2005, Hirohata:2005,Tezuka:2006,Takahashi:2007, Schebaum:2010}), much less is known about the magnetic microstructure and its impact on the magnetotransport properties of Heusler thin films. This is all the more surprising as the modelling of magnetoresistive effects such as TMR, giant magnetoresistance~\cite{Baibich:1988,Binasch:1989}, anisotropic magnetoresistance~\cite{Thomson:1857} or the angle dependent magnetoresistance~\cite{Limmer:2006} are usually based on the assumption of a macrospin, i.e., a single domain state.

\begin{figure}
  \includegraphics{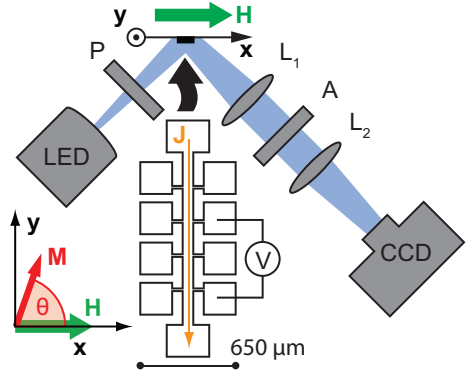}\\
  \caption{(Color online) Sketch of experimental setup and sample. Linearly s-polarized light ($\lambda=\unit{455}{\nano\meter}$) impinges on the sample at an angle of incidence of 45\degree. Using two lenses $L_1$ and $L_2$ with focal length $f=\unit{150}{\milli\meter}$ and the polarization analyzer $A$, the spatially resolved MOKE signal is recorded with a CCD-camera. The external magnetic field $\mathbf{H}$ is applied along the $\mathbf{x}$ axis orthogonal to the current $\mathbf{J}$. The magnetoresistance is recorded in a four point measurement between the indicated contact pads. $\Theta$ denotes the magnetization $\mathbf{M}$ orientation with respect to the $\mathbf{x}$ axis.}\label{fig:setup}
\end{figure}

In this letter, we report on simultaneous magnetotransport and spatially resolved magneto optical Kerr effect (MOKE) measurements in Heusler thin films at room temperature. We use Co$_2$FeAl as a prototype Heusler material, as thin films with state-of-the-art structural and magnetic properties can be deposited directly onto MgO substrates. The thin films were prepared by DC- and RF-sputtering on a MgO (001) substrate at a base pressure of $\unit{1\times10^{-7}}{mbar}$. We here focus on a sample consisting of \unit{5}{\nano\meter} MgO, \unit{50}{\nano\meter} Co$_2$FeAl and \unit{1.8}{\nano\meter} MgO annealed for one hour at \unit{500}{\celsius}. The sample was patterned into the Hall bar geometry shown in Fig.~\ref{fig:setup} by optical lithography and Ar ion beam etching. Using angle dependent magnetoresistance~\cite{Limmer:2006} at room temperature, we determined that the sample shows dominantly cubic magnetic anisotropy in the film plane with easy axes (e.a.) along the crystalline $[110]$ and $[\bar{1}10]$ directions which are parallel and perpendicular to the main Hall bar (along $\mathbf{y}$ and $\mathbf{x}$), respectively.

\begin{figure}
  \includegraphics{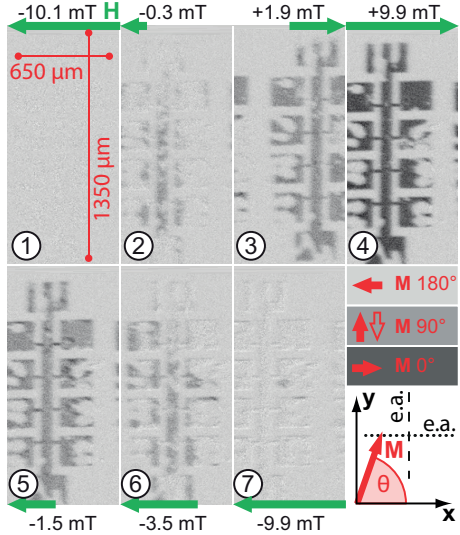}\\
  \caption{(Color online) Selected spatially resolved MOKE images. Images 1 to 4 were taken during the magnetic field upsweep and images 5 to 7 during the downsweep at the indicated values of $\mu_0H$. The $\mathbf{H}$ orientation is shown by the green arrows.  The shading represents the $\mathbf{M}$-orientation in each domain, as indicated in the lower right. We observe domains with $\mathbf{M}$ orientated along the e.a.\ along $\mathbf{y}$ (dashed) and the e.a.\ along $\mathbf{x}$ (dotted).}\label{fig:domains}
\end{figure}

\begin{figure}
  \includegraphics{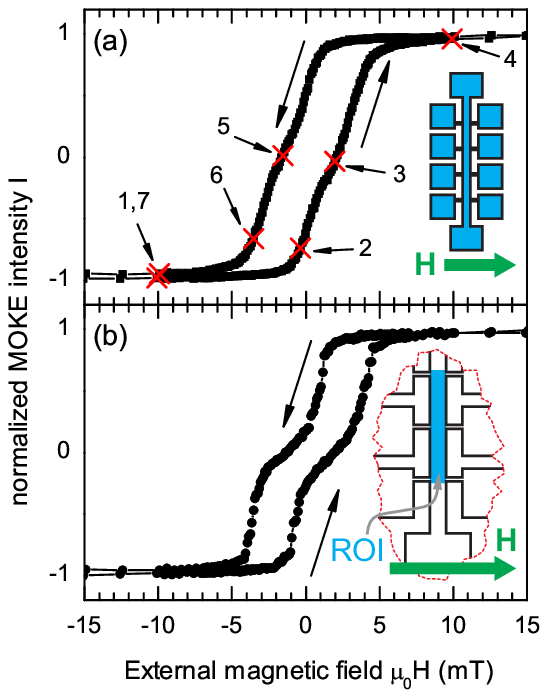}\\
  \caption{(Color online) (a) MOKE intensity $I$ integrated over the entire Hall bar and normalized to $[-1,1]$ as a function of $\mu_0H$. The numbers correspond to the MOKE images shown in Fig.~\ref{fig:domains}. (b) Normalized $I(H)$ integrated over the indicated region of interest (ROI). The two-step magnetic switching characteristic of cubic magnetic anisotropy is clearly evident from the data.}\label{fig:MOKE}
\end{figure}

The MOKE images shown in Fig.~\ref{fig:domains} were recorded in the longitudinal MOKE configuration~\cite{Kerr:1877, Yang:1993} as schematically shown in Fig.~\ref{fig:setup}. S-polarized light is used to illuminate the sample and the p-polarized component of the reflected light is imaged onto a Andor Luca-S CCD using two lenses. This simple MOKE setup allows for lateral spatial resolution of approximately \unit{10}{\micro\meter}. Prior to image acquisition, we prepared the sample in a magnetically saturated state by applying $\mu_0 H=\unit{-30}{\milli\tesla}$ along $\mathbf{x}$ and iteratively adjusted the analyzer $A$ and polarizer $P$ to obtain minimal total intensity on the CCD. The analyzer was subsequently rotated 1\degree out of extinction for the measurement. Sweeping the magnetic field up to $\mu_0 H=\unit{+30}{\milli\tesla}$ and back to $\mu_0 H=\unit{-30}{\milli\tesla}$ (in steps of \unit{0.1}{\milli\tesla} for $\mu_0|H|<\unit{10}{\milli\tesla}$), MOKE images were recorded at each field point. To obtain magnetic contrast, a reference image recorded in saturation is subtracted from each image. A selection of the resulting difference images is displayed in Fig.~\ref{fig:domains}. At \unit{-10.1}{\milli\tesla} [image (1)] the sample is still in the magnetically saturated single-domain state so that no magnetic contrast is visible. Upon increasing the external magnetic field strength, domains start to nucleate and propagate [images (2) and (3)]. In image (3) most parts of the Hall bar show identical grey shading, corresponding to $\mathbf{M}$ along $\mathbf{y}$, except for the aluminum bond wires visible as white patches on the contact pads. By further increasing $\mu_0 H$, the magnetic contrast can be increased once again as visible in the change of the Hall bar shading from grey in image (3) to dark grey in image (4). Dark grey hereby corresponds to $\mathbf{M}\parallel\mathbf{x}$. This two-step magnetic switching behavior (antiparallel, perpendicular, parallel to $\mathbf{x}$, as indicated in the lower right of Fig.~\ref{fig:domains}) is characteristic for cubic anisotropy~\cite{Cowburn:1995}. Similar domain states are observed during the magnetic field downsweep [images (5) to (7)].
Figure~\ref{fig:MOKE}(a) shows the normalized MOKE intensity obtained upon integrating the MOKE signal within the entire Hall bar region.  It shows the two-step shape characteristic for cubic magnetic anisotropy and allows to directly compare the spatially resolved domain contrast shown in Fig.~\ref{fig:domains} to the integral magnetic contrast in Fig.~\ref{fig:MOKE}(a) at the highlighted data points marked with the image numbers.
For comparison of our MOKE and magnetotransport data, we also integrated the MOKE intensity in a region of interest (ROI) corresponding to the region probed by magnetotransport. The resulting $I(H)$ is shown in Fig.~\ref{fig:MOKE}(b). It again clearly exhibits the dual switching behavior indicative of cubic magnetic anisotropy.
\begin{figure}
  \includegraphics{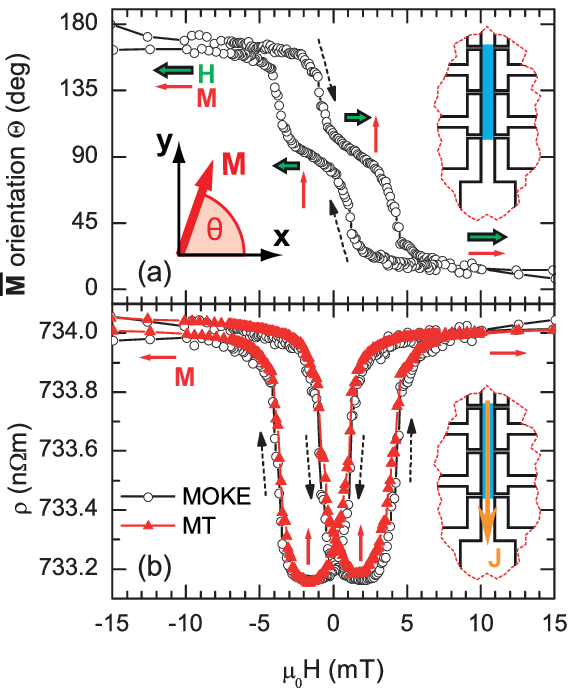}\\
  \caption{(Color online) (a) \Mm orientation $\Theta(H)$ calculated from the integral MOKE intensity in the indicated region of interest. (b) The AMR calculated from $\Theta(H)$ perfectly traces the experimentally observed $R(H)$.}\label{fig:AMR}
\end{figure}
To extract an effective, average magnetization direction from the MOKE images, we calculate the mean magnetization direction \Mm in the ROI as a function of $\mu_0 H$, assuming that \Mm is parallel to $\mathbf{H}$ for $\mu_0 |H|=\unit{30}{\milli\tesla}$. More precisely, we take the \Mm orientation to be $\Theta_-=180\degree$ for $\mu_0 H=\unit{-30}{\milli\tesla}$ and $\Theta_+=0\degree$ for $\mu_0 H=\unit{+30}{\milli\tesla}$. This is a valid assumption as the integral MOKE loops displayed in Fig.~\ref{fig:MOKE} exhibit hysteresis closure at approx. $\unit{\pm5}{\milli\tesla}$. With the normalized MOKE intensity $I(H)$ in the ROI as shown in Fig.~\ref{fig:MOKE}(b), we can now calculate $\Theta(H)$ as~\cite{Gil:2005}
\begin{equation}\label{eq:Theta}
    \Theta(H)=\arccos\left[I(H)\right]\;.
\end{equation}
The resulting $\Theta(H)$ is shown in Fig.~\ref{fig:AMR}(a). The double switching behavior is again clearly visible, with $\Theta\approx90\degree$ at small absolute values of $\mu_0H$. Hence, we observe \Mm switching from $\Mm \parallel -\mathbf{x}$ (first e.a., parallel to $\mathbf{H}$) to $\Mm \parallel \mathbf{y}$ (second e.a., perpendicular to $\mathbf{H}$) and subsequent switching to $\Mm \parallel \mathbf{H}$ again. Note that the L-MOKE measurement geometry is sensitive only to the projection of \Mm on $\mathbf{H}\parallel \mathbf{x}$, therefore it is not possible to discriminate between the energetically degenerate \Mm orientations $\Theta=90\degree$ and $\Theta=270\degree$.

Having determined $\Theta(H)$ we can now calculate the magnetoresistance $\rho(H)$ expected in the macrospin model and compare it to four point longitudinal magnetotransport data acquired simultaneously to the MOKE images. The magnetotransport measurements were carried out with the contact geometry sketched in Fig.~\ref{fig:setup} and a current $J=\unit{5}{\milli\ampere}$. The results are shown by the red triangles in Fig.~\ref{fig:AMR}(b). The resistivity changes from $\rho_\perp=\unit{734.1}{\nano\ohm\meter}$ at $\mu_0H=\unit{-30}{\milli\tesla}$ (negative saturation, $\mathbf{M} \parallel -\mathbf{x}$) to $\rho_\parallel=\unit{733.2}{\nano\ohm\meter}$ at $\mu_0H=\unit{2}{\milli\tesla}$ ($\mathbf{M} \parallel \mathbf{y}$) in the magnetic field upsweep. In the following, we take  $\rho_\perp$ and $\rho_\parallel$ as the resistivity for \Mm perpendicular and parallel to $\mathbf{J}$, respectively. The anisotropic magnetoresistance (AMR) $(\rho_\perp-\rho_\parallel)/(\frac{1}{2}(\rho_\perp+\rho_\parallel))\approx1.2\times10^{-3}$ compares well to the value reported for Co$_2$MnGe Heusler compounds~\cite{Ambrose:2000}.

We now calculate the AMR from the effective macrospin \Mm [cf. Fig.~\ref{fig:AMR}(a)] using~\cite{McGuire:1975}
\begin{equation}\label{eq:AMR}
   \rho(H)=\rho_\perp +(\rho_\parallel-\rho_\perp)\cos^2\left[\Theta(H)+\Phi\right]\;,
\end{equation}
where $\Phi=270\degree$ is the angle between the current direction $\mathbf{J}$ and the $\mathbf{x}$-axis. The result is depicted by the open circles in Fig.~\ref{fig:AMR}(b) and shows excellent agreement with the AMR determined by magnetotransport measurements.  This shows that, in Co$_2$FeAl Heusler compounds, it is possible to model the AMR using a simple macrospin model that neglects the domain wall resistance, although microscopically a complex domain pattern is observed (cf. Fig.~\ref{fig:domains}).

In conclusion, we compared magnetic microstructure and magnetotransport properties in Co$_2$FeAl Heusler compounds by simultaneously recording spatially resolved magneto optical Kerr effect and magnetotransport data. An effective magnetization orientation (macrospin) corresponding to the spatially averaged microscopic $\mathbf{M}$ configuration in the region probed by magnetotransport was extracted from the MOKE images. We found that the magnetotransport properties can be quantitatively reproduced assuming that this macrospin determines the magnetoresistance. This demonstrates that even if the investigated Heusler microstructure exhibits a complex magnetic domain pattern, a macrospin model fully suffices to describe its magnetotransport properties. Hence, the contributions of domain walls to the magnetoresistance are negligible. This opens the path for further investigations of Heusler compound thin films, using macrospin-based magnetotransport techniques.

Financial support via DFG Project No.~GO 944/3-1 and the German Excellence Initiative via the 'Nanosystems Initiative Munich (NIM)' is gratefully acknowledged. I.-M.I.\ and A.T.\ are supported by a MIWF junior researcher grant.


\begin{thebibliography}{10}

\bibitem{Kandpal:2007}
H.~C. Kandpal, G.~H. Fecher, and C.~Felser, J. Phys. D: Appl. Phys.
  \textbf{40}, 1507 (2007).

\bibitem{Trudel:2010}
S.~Trudel, O.~Gaier, J.~Hamrle, and B.~Hillebrands, J. Phys. D: Appl. Phys.
  \textbf{43}, 193001 (2010).

\bibitem{Julliere:1975}
M.~Julliere, Phys. Lett. A \textbf{54}, 225  (1975).

\bibitem{Ishikawa:2009}
T.~Ishikawa, H.-x. Liu, T.~Taira, K.-i. Matsuda, T.~Uemura, and M.~Yamamoto,
  Appl. Phys. Lett. \textbf{95}, 232512 (2009).

\bibitem{Ebke:2009}
D.~Ebke, V.~Drewello, M.~Sch\"{a}fers, G.~Reiss, and A.~Thomas, Appl. Phys.
  Lett. \textbf{95}, 232510 (2009).

\bibitem{Inomata:2006}
K.~Inomata, S.~Okamura, A.~Miyazaki, M.~Kikuchi, N.~Tezuka, M.~Wojcik, and
  E.~Jedryka, J. Phys. D: Appl. Phys. \textbf{39}, 816 (2006).

\bibitem{Hirohata:2005}
A.~Hirohata, H.~Kurebayashi, S.~Okamura, T.~Masaki, T.~Nozaki, M.~Kikuchi,
  N.~Tezuka, K.~Inomata, J.~S. Claydon, and Y.~B. Xu, J. Appl. Phys.
  \textbf{97}, 10C308 (2005).

\bibitem{Tezuka:2006}
N.~Tezuka, S.~Okamura, A.~Miyazaki, M.~Kikuchi, and K.~Inomata, J. Appl. Phys.
  \textbf{99}, 08T314 (2006).

\bibitem{Takahashi:2007}
Y.~Takahashi, T.~Ohkubo, K.~Hono, S.~Okamura, N.~Tezuka, and K.~Inomata, J.
  Magn. Magn. Mater. \textbf{313}, 378  (2007).

\bibitem{Schebaum:2010}
O.~Schebaum, D.~Ebke, A.~Niemeyer, G.~Reiss, J.~S. Moodera, and A.~Thomas, J.
  Appl. Phys. \textbf{107}, 09C717 (2010).

\bibitem{Baibich:1988}
M.~N. Baibich, J.~M. Broto, A.~Fert, F.~N. Van~Dau, F.~Petroff, P.~Etienne,
  G.~Creuzet, A.~Friederich, and J.~Chazelas, Phys. Rev. Lett. \textbf{61},
  2472 (1988).

\bibitem{Binasch:1989}
G.~Binasch, P.~Gr\"unberg, F.~Saurenbach, and W.~Zinn, Phys. Rev. B
  \textbf{39}, 4828 (1989).

\bibitem{Thomson:1857}
W.~Thomson, Proc. R. Soc. Lond. \textbf{8}, 546 (1857).

\bibitem{Limmer:2006}
W.~Limmer, M.~Glunk, J.~Daeubler, T.~Hummel, W.~Schoch, R.~Sauer, C.~Bihler,
  H.~Huebl, M.~S. Brandt, and S.~T.~B. Goennenwein, Phys. Rev. B \textbf{74},
  205205 (2006).

\bibitem{Kerr:1877}
J.~Kerr, Philos. Mag. \textbf{3}, 321 (1877).

\bibitem{Yang:1993}
Z.~J. Yang and M.~R. Scheinfein, J. Appl. Phys. \textbf{74}, 6810 (1993).

\bibitem{Cowburn:1995}
R.~P. Cowburn, S.~J. Gray, J.~Ferr\'{e}, J.~A.~C. Bland, and J.~Miltat, J.
  Appl. Phys. \textbf{78}, 7210 (1995).

\bibitem{Gil:2005}
W.~Gil, D.~G\"{o}rlitz, M.~Horisberger, and J.~K\"{o}tzler, Phys. Rev. B
  \textbf{72}, 134401 (2005).

\bibitem{Ambrose:2000}
T.~Ambrose, J.~J. Krebs, and G.~A. Prinz, J. Appl. Phys. \textbf{87}, 5463
  (2000).

\bibitem{McGuire:1975}
T.~McGuire and R.~Potter, IEEE Trans. Magn. \textbf{11}, 1018  (1975).

\end{thebibliography}
\end{document}